\documentstyle[emulateapj]{article}

\lefthead{Armitage, Reynolds \& Chiang}
\righthead{Accretion flows crossing the last stable orbit}

\submitted{ApJ, in press}

\begin{document}

\title{Simulations of accretion flows crossing the last stable orbit}
        
\author{Philip J. Armitage\altaffilmark{1,2}}
\affil{Max-Planck-Institut f\"ur Astrophysik, 
        Karl-Schwarzschild-Str. 1, \\ D-85741 Garching, Germany \\
        {\tt pja3@st-andrews.ac.uk}}
\author{Christopher S. Reynolds\altaffilmark{3} and James Chiang}
\affil{JILA, Campus Box 440, University of Colorado, Boulder CO 80303, 
        USA \\
        {\tt chris@rocinante.colorado.edu; chiangj@rocinante.colorado.edu}}           
        
\altaffiltext{1}{CITA, University of Toronto, McLennan 
        Labs, 60 St George Street, Toronto, Ontario M5S 3H8, Canada} 
 
\altaffiltext{2}{Present address: School of Physics and Astronomy, 
       	University of St Andrews, North Haugh, St Andrews, Fife, 
	KY16 9SS, UK}
               
\altaffiltext{3}{Hubble Fellow}

\begin{abstract}
  We use three dimensional magnetohydrodynamic simulations, in a
  pseudo-Newtonian potential, to study geometrically thin accretion disc 
  flows crossing $r_{\rm ms}$, the marginally stable circular orbit around black 
  holes. We concentrate on vertically unstratified and isothermal disk models, but 
  also consider a model that includes stratification. In all cases, we 
  find that the sonic point lies just inside $r_{\rm ms}$, with a
  modest increase in the importance of magnetic field energy, relative 
  to the thermal energy, observed inside the last stable orbit.
  The time-averaged gradient of the specific angular momentum of the flow, $(dl/dr)$, is 
  close to zero within $r_{\rm ms}$, despite the presence of large 
  fluctuations and continuing 
  magnetic stress in the plunging region. The result that the 
  specific angular momentum is constant within $r_{\rm ms}$ is in 
  general agreement with traditional disk models computed using a 
  zero-torque boundary condition at the last stable orbit.
\end{abstract}


\section{Introduction}
The existence of a marginally stable orbit is a distinctive 
feature of accretion flows onto black holes. Within $r_{\rm ms}$, stable 
circular orbits do not exist, and gas inevitably plunges on a short 
timescale into the hole. The location of the marginally stable orbit, 
which lies at $r_{\rm ms} = 6 GM / c^2$ for a non-rotating hole, varies 
strongly with the spin parameter $a$ for Kerr black holes. This 
variation, together with the assumption of a large change in the emission properties 
of the gas interior and exterior to $r_{\rm ms}$, is central to 
all attempts to measure $a$ from observable quantities for both 
galactic (Zhang, Cui \& Chen 1997) and supermassive black holes 
(Iwasawa et al. 1996; Dabrowski et al. 1997; Bromley, Chen \& 
Miller 1997). Neutron stars, if they are sufficiently compact, 
could also possess a last stable orbit, though we will not 
consider this possibility further here.

The presence of gas on plunging orbits within $r_{\rm ms}$ can have observable 
consequences. It potentially modifies the profile of the relativistic 
iron K$\alpha$ line that has been observed in the X-ray spectra 
of some Seyfert galaxies (Tanaka et al. 1995; Reynolds \& 
Begelman 1997; Young, Ross \& Fabian 1998), and may create 
detectable absorption signatures (Nandra et al. 1999). However, 
it has generally been believed that gas within $r_{\rm ms}$ 
does not have any {\em dynamical} importance, since the infall 
rapidly becomes supersonic a short distance inside the marginally 
stable orbit. Whatever complexities may develop in the flow 
subsequently are then causally disconnected from the disk at larger radius. 
This line of reasoning suggests that for modelling the disk (by 
which we mean the region of the flow outside the last stable orbit), 
it suffices to impose a zero-torque boundary condition at 
$r_{\rm ms}$ and ignore the interior region altogether. 
For the details of disk models constructed in this manner, 
we refer the reader to Abramowicz \& Kato (1989), and references therein.

This simple hydrodynamic picture would be oversimplified if magnetic fields inside
$r_{\rm ms}$ were strong enough to maintain a connection 
between the plunging region and the disk. A recent analysis 
by Krolik (1999) showed that if magnetic fields, generated 
in the disk, remain frozen into the gas as it inspirals 
within $r_{\rm ms}$, then the energy density in the fields 
can become comparable to the rest-mass energy density of 
the flow. Related ideas have also been proposed by Gammie (1999). 
In addition to altering the zero-torque boundary condition for the 
disk, the presence of extremely strong fields in the plunging 
region could have several potentially observable consequences (Agol \& 
Krolik 2000). For example, the radiative efficiency of the disk, 
$\epsilon = L / {\dot M} c^2$, could be substantially increased. 
The analysis of Krolik (1999), however, depends upon a split of the accretion 
flow into two distinct regions, a disk at $r > r_{\rm ms}$ in 
which magnetic dynamo processes (Balbus \& Hawley 1991; Brandenburg et al. 1995; 
Stone et al. 1996; for a review see e.g. Hawley \& Balbus 1999) 
maintain a turbulent state, and a 
plunging region at $r < r_{\rm ms}$ in which it is assumed that magnetic 
flux is frozen into the fluid. This split is clearly an 
approximation, because the same turbulent processes that act to 
generate and destroy magnetic fields in the disk itself will continue to 
operate until some finite distance inside the last stable 
orbit. This may alter the conclusion that a rapid 
growth in the relative importance of magnetic fields 
is inevitable (Paczynski 2000). The analytic analysis further assumes that 
the flow is axisymmetric and steady-state. This is also only approximately
the case in a turbulent accretion flow, and needs to be tested via numerical 
simulations.

In this paper, we employ three dimensional magnetohydrodynamic (MHD) 
simulations to study the properties of the accretion flow as 
it crosses $r_{\rm ms}$. A pseudo-Newtonian potential 
appropriate for a non-rotating black hole (Paczynski \& 
Wiita 1980) is used to mimic the relativistic effect of a last 
stable orbit within a non-relativistic MHD code. Within this 
approximation, simulating the evolution of magnetic fields 
within the plunging region is, if anything, expected to be 
easier than following the development of turbulence in the 
disk at larger radii. Within the disk, orbits are 
stable, and thus in the absence of turbulence a field loop 
in the plane of the disk will {\em always} be sheared out 
in azimuth until numerical reconnection sets in at the 
grid scale. Once into the plunging region there is only a 
finite (but large) amount of shear before the infalling gas 
reaches the black hole.

The accretion flow within $r_{\rm ms}$ has already been 
studied by Hawley (2000), who presented global MHD simulations of accretion 
tori with a geometry similar to that of popular ADAF (Narayan \& 
Yi 1994) and ADIOS (Blandford \& Begelman 1999) models for 
radiatively inefficient flows. This geometry is thought to be 
appropriate at relatively low accretion rates for both 
supermassive and stellar mass black hole systems (for 
observational support see, e.g., Gilfanov, Churazov \& Revnivtsev 1999).
The specific model considered was geometrically fairly 
thick at large radius ($r \gg r_{\rm ms}$), although near 
the last stable orbit the importance of pressure forces, 
and the vertical scale height, was smaller. Indeed, at 
$r = r_{\rm ms}$, the relative scale height was $(h/r) \sim 0.1$, 
which is similar to the value used in our simulations. 
Significant magnetic stresses 
within $r_{\rm ms}$ were obtained in these simulations, and in 
subsequent higher resolution simulations of the same geometry
(Hawley \& Krolik 2000). 

Existing work suggests that the 
conditions at $r_{\rm ms}$ may differ for geometrically thin 
and thick disks (Popham \& Gammie 1998). We 
therefore emphasize that in this 
paper we address only the 
case of geometrically thin disks. The existence of a thin 
disc implies that radiative cooling must be efficient, and 
this allows us to further simplify the problem by adopting 
an isothermal equation of state in lieu of solving the energy 
equation.
Thin disks are the expected mode 
of accretion at high $\dot{M}$ in both Active Galactic Nuclei, 
and in stellar mass Galactic black hole sources. 

\section{Numerical methods}

\subsection{The ZEUS hydrodynamics code}

We simulate the accretion flow using the ZEUS 
code (Stone \& Norman 1992a,1992b; Clarke, Norman \& Fielder 1994; 
Norman 2000) to solve the equations of ideal magnetohydrodynamics.
ZEUS is an explicit Eulerian finite difference code, formally 
of second order accuracy, which uses an artificial viscosity to 
reproduce shocks. Advantages of the code for accretion disk 
applications include a flexible choice of gridding and co-ordinate 
systems, and algorithms (Norman, Wilson \& Barton 1980) that minimise 
spurious diffusion of angular momentum relative to mass.

The analyis of the behavior of magnetic fields in the plunging region 
by Krolik (1999) is independent of the existence of vertical 
stratification in the disk. For our initial simulations we therefore
adopt the computationally easiest option, and consider a 
vertically unstratified disk (i.e. one where the vertical component 
of gravity is artificially set to zero), and an 
isothermal fluid, where the presure $P$ is given by,
\begin{equation} 
 P = \rho c_s^2,
\label{eq1}
\end{equation} 
where $c_s$ is the sound speed and $\rho$ is the density. 
For our standard model we choose $c_s$ such that the 
ratio of the sound speed to the circular orbital velocity at the 
last stable orbit, $c_s / v_\phi = 0.08$. In an unstratified 
simulation, of course, $c_s$ plays no role in setting the 
vertical scale height of the flow. However, the ratio 
$c_s / v_\phi$, which in a real disk is related to the 
relative disk scale height via $h/r \simeq c_s / v_\phi$, 
still determines the importance of radial pressure 
forces for the disk structure. The simulated flows 
are `thin' in the sense that radial pressure forces, which scale 
as $(h/r)^2$, are small compared to the gravitational force. 
A model with smaller sound speed was also investigated. For better comparison 
with the simulations of Hawley (2000), we also ran a variant 
including the effect of vertical stratification.  

Standard choices for the Courant number, $C_0 = 0.5$, and coefficient of 
artificial viscosity, $C_2 = 2.0$, were used throughout. Second order 
advection was used for all quantities.
The only code scaling which needs to be mentioned is that 
for time. We adopt units in which the orbital period at $r_{\rm ms}$, 
$P_{\rm ms} = 7.7$.

\subsection{Paczynski-Wiita potential}

ZEUS is a Newtonian fluid dynamics code which does not model 
any of the effects of special or general relativity. Within 
this framework, we use a pseudo-Newtonian gravitational potential
(Paczynski \& Wiita 1980), 
\begin{equation} 
 \psi = - { {GM} \over {r - r_g} } 
\label{eq_pw}
\end{equation}
where $r_g = 2 GM / c^2$, to model what is expected to be the 
dominant relativistic effect around a non-rotating black hole -- 
the existence of an innermost stable orbit at $r_{\rm ms} = 6 GM / c^2$. 

\subsection{Initial and boundary conditions}

We simulate a wedge of the disk extending $30^\circ$ in azimuth 
in cylindrical polar geometry, $(r,z,\phi)$. This limitation on the 
azimuthal extent of the domain has the undesirable effect of 
eliminating larger scale azimuthal modes of the magnetic field, which 
contribute significantly to the total power in global disk 
simulations (e.g. Armitage 1998). The local properties 
of disk turbulence (for example the relative strengths of 
radial and azimuthal field components), however, 
which are probably of more importance for this 
problem, are less affected by the size of the azimuthal domain.

The boundary conditions are periodic in $\phi$, and set to outflow 
at both $r_{\rm min}$ and $r_{\rm max}$. Outflow boundary conditions 
are implemented in ZEUS by setting zero gradients for all flow 
variables at the boundary. We note that this choice of boundary 
conditions allows a significant toroidal magnetic field to be 
advected across the inner boundary -- we have not attempted 
to impose Newtonian analogs of the general relativistic
boundary conditions for magnetic fields at the event horizon 
of the black hole. From a numerical standpoint, the use of outflow 
boundary conditions is 
desirable to ensure that the simulations do indeed model magnetic 
instabilities, rather than purely hydrodynamic instabilities to 
which relatively narrow annuli, with reflecting boundaries, are 
known to be susceptible (e.g. Narayan \& Goodman 1989). We use 
a uniform grid in both $\phi$ and $z$. In the radial direction, 
a uniform grid interior to $r_{\rm ms}$ is matched smoothly 
onto a grid at larger radii for which $\Delta r_{i+1} = k \Delta r_{i}$, 
with $k > 1$ a constant. This concentrates resolution in the 
region of greatest interest near, and within, the marginally stable orbit.

The initial conditions for the calculation are intended as thin 
disk analogs of the toroidal configuration used by Hawley (2000), 
with the additional assumption of isothermality. 
We take a gaussian surface density profile, centered at $r=2 r_{\rm ms}$, 
with an azimuthal velocity appropriate to Keplerian rotation. We 
relax this profile in a preliminary non-magnetic one dimensional calculation, to 
ensure that it is an accurate numerical equilibrium state. The 
resulting profile of $\rho$ and $v_\phi$ is then mapped into three 
dimensions, and a seed magnetic field added. To evaluate the 
sensitivity of the results to the boundary conditions, three unstratified 
simulations were run, with different 
choices of seed field and associated boundary conditions in $z$. 
Two simulations were run with an initially vertical field 
imposed at all radii where the surface density exceeded a 
threshold value, taken to be approximately a quarter of the 
maximum value of the surface density.
Periodic boundary conditions for all variables were imposed in the $z$ direction. 
For the first simulation (the `standard' run) the radial and vertical 
boundaries were at $r_{\rm min} = 1$, 
$r_{\rm max} = 5$, and $z = \pm 0.3$, in code units where $r_{\rm ms} = 1.5$.
The initial field strength (in regions seeded with field) was such that 
the ratio of gas pressure to magnetic pressure, $\beta_z = 5000$. The numerical 
resolution was $n_r = 150$, $n_z = 32$ and $n_\phi = 40$ grid points,
with 30 of the radial grid points interior to $r_{\rm ms}$. The second run simulated a
cooler disk (with a sound speed half the previous value), in a larger 
computational volume, bounded by $1 < r < 10$ and $z = \pm 0.25$. This run 
used $n_r = 210$, $n_z = 32$ and $n_\phi = 48$ grid points, and an 
initial seed field $\beta_z = 800$. 
Finally, a simulation was run with parameters similar to the first, 
except starting with an initially azimuthal field, 
with $\beta_\phi = 100$. For this simulation the vertical boundary 
conditions were chosen to be reflecting, with the vertical 
components of both the velocity and the magnetic field set to 
zero on the boundaries. The resolution for this run 
was $120 \times 32 \times 40$ 
grid points.

Simulations that include vertical stratification are obviously 
more realistic, but they are also more demanding 
of computational resources, because the development of magnetically 
dominated low density regions at high $\vert z \vert$ places severe 
restrictions on the timestep. Numerical tricks can be used to mitigate 
this problem (Miller \& Stone 2000), but for this paper we adopted the 
simpler approach of considering a volume containing only a small number 
of disk scale heights. We therefore reran the standard model, including 
the vertical component of gravity, in a domain bounded by 
$z = \pm 0.5$. This admits only a couple of scale heights at 
$r_{\rm ms}$. As in the standard model, periodic boundary 
conditions were used for the magnetic field in the vertical 
direction. For this run we used $n_r = 150$, $n_z = 48$ and 
$n_\phi = 40$ grid points.

It is worth noting at the outset that these simulations are not, 
and are not intended to be, realizations of the same physical 
situation, and thus the results will differ between them. An 
initially vertical field gives the fastest possible growth rate of 
the Balbus-Hawley (1991) instability. Additionally, a non-zero vertical flux 
boosts the strength of MHD turbulence and increases 
the effective Shakura-Sunyaev (1973) $\alpha$ 
parameter obtained (Hawley, Gammie \& Balbus 1995). Substantially 
longer timescales are required to reach saturation with an 
initially azimuthal field. We have run both simulations that 
used the smaller computational domain until 
a significant fraction (around a half) of the initial mass had been 
accreted, and compare the results at this late stage when the 
flow in the inner regions of the disk, near $r_{\rm ms}$, has 
reached an approximate steady state.  

\section{Results}

Fig.~1 shows the growth in the magnetic energy of the radial magnetic 
field component in the standard unstratified model with an initially 
vertical ($z$ direction, $\beta_z = 5000$) 
magnetic field. Around 10 orbits of evolution are 
required to reach a nonlinear state in the inner regions of the 
flow, during which time the radial magnetic field energy grows 
exponentially. The growth rate is consistent with the expected 
growth rate of the Balbus-Hawley instability in this field 
geometry (Balbus \& Hawley 1991), evaluated at the smallest radius 
that has been seeded with magnetic field. Subsequent to the instabilities 
saturating, the field energy fluctuates but remains roughly 
constant, before finally starting to decline as a significant 
fraction of the disk is accreted. At the point the run was stopped, 
at $t=200$, 65\% of the initial mass had been accreted. The magnetic 
field was dominated by the toroidal component, with the ratio of 
magnetic field energy in the $z$, $r$ and $\phi$ fields being 
approximately 1:2:30.

The simulation seeded with an azimuthal magnetic field shows similar 
behavior, but with much slower growth of the magnetic field energy, 
despite the initially higher ratio of the magnetic energy to the 
thermal energy. This configuration was run up to $t=600$,
by which time just under 40\% of the initial disk mass 
had been accreted.

Fig.~2 shows the evolution of the disk surface density with time, 
for the standard run with an initial vertical field. The evolution differs 
somewhat from the standard diffusive evolution of a viscous 
annulus (e.g. Pringle 1981), due to both the varying amount of time required 
before the disk at different radii becomes fully turbulent, and 
due to the development of supersonic infall interior to $r_{\rm ms}$. 
However the general trend towards a broadening surface density 
profile, that remains relatively smooth, is recovered. Spatial 
fluctuations in $\Sigma$, shown in Fig.~3, are strongly sheared and 
thus predominantly azimuthal in extent. When normalised to the 
mean surface density as a function of radius, as in the figure, there 
is no obvious visual indication of the location of the marginally stable 
orbit.

The radial and azimuthal velocities as a function of radius obtained in 
the standard run are shown in Fig.~4. A small radial velocity within the 
disk itself transitions smoothly into rapid infall within the 
marginally stable orbit. The sonic point lies somewhat inside $r_{\rm ms}$ 
(within a radial distance $\sim h$, the disk scale height that would 
correspond to the assumed sound speed). Note that the radial velocity 
slightly {\em outside} 
the marginally stable orbit is already beginning to increase in 
magnitude, in advance of reaching the actual infall region.
Beyond $r \approx 2 r_{\rm ms}$, the radial velocity is {\em outward}. 
This is a consequence of the initial surface density profile and the 
use of a free outer boundary condition. We have also experimented 
with runs in which mass was continually injected across the outer 
boundary, and the disk allowed to evolve until a quasi-steady state 
was achieved. These runs yielded similar results to those 
reported here -- in particular there was no strong amplification of 
magnetic field observed interior to $r_{\rm ms}$. However, with 
inflow boundary conditions at $r_{\rm out}$ it is hard to be sure 
that purely hydrodynamic instabilities, which are known to occur within 
relatively narrow annuli when the boundary conditions are reflecting, 
do not contaminate the results. We therefore concentrate attention 
on the current simulations which follow a ring of gas which is allowed both 
to accrete and to spread to larger radii.

As expected for a thin disk, in which radial pressure forces are 
small compared to the gravitational force, the azimuthal velocity 
in the disk, $v_\phi$, is very close to the Keplerian value for 
orbits in a Paczynski-Witta potential.

The presence of magnetic fields in the flow potentially allows 
the flow interior to the sonic point to communicate with the 
exterior disk. Generically, except in unusual circumstances, 
MHD turbulence driven by the Balbus-Hawley instability saturates 
at a level where the magnetic pressure in the disk is substantially less 
than the thermal pressure, typically by one or two orders of 
magnitude (Hawley, Gammie \& Balbus 1995; Brandenburg et al. 1995; 
Stone et al. 1996). Stronger fields, relative to the gas pressure, 
are likely to occur in the disk corona (Miller \& Stone 2000).
As shown in Fig.~5 for our standard unstratified 
simulation, the mean Alfven speed in the disk, $v_A = \sqrt{(B^2 / 4 \pi \rho)}$, 
remains less than the sound speed at all radii. This is even more 
true for the simulation with the initial $\phi$ field, not shown 
in the figure, since the saturation level of 
the magnetic fields in that run is significantly below the value 
obtained in the initial vertical field runs. Similarly, for the vertically 
stratified version of the standard run, the mean Alfven speed is smaller 
than the sound speed at all radii, and is of similar magnitude to the 
unstratified case. For this run, the magnetic fields and the Alfven 
speed near the disk midplane, also plotted in Fig.~5, are significantly 
greater than the mean value in the plunging region, though not 
by a large factor.

These results do not, however, exclude the possibility that the plunging region 
can exert a torque on the disk at the last stable orbit.
In the presence of turbulence, some regions of the 
flow have relatively stronger magnetic fields (or lower density), 
with correspondingly higher Alfven speed. As a result, the 
critical point can penetrate further into the plunging 
region  than one would estimate based solely on the mean 
flow properties. As shown in Fig.~5, for both the stratified 
and the unstratifed runs, we find that the {\em peak} Alfven speed at 
any radius is substantially greater than the mean value. Although 
it does not rise as rapidly as the absolute value of the radial 
velocity, it is therefore possible, at least in principle, 
for the plunging region some distance inside the last stable 
orbit to be in communication with the disk at larger radii.
In our simulations this distance is still only $\sim h$, but 
whether this result also holds for cooler flows is currently unknown.

Fig.~6 shows the ratio of the magnetic energy to the thermal energy 
for the standard run, and for the run with an 
initially azimuthal seed magnetic field, as a function of radius. A single timeslice from the 
simulation produces a rather noisy estimate of this quantity, so 
we plot the average from 5 independent epochs close to the end of 
the run ($t=150$ to $t=200$ for the $z$ field run, and
$t=440$ to $t=600$ 
for the $\phi$ field case). The ratio $(B^2 / 8 \pi \rho c_s^2)$  
is around 0.05 - 0.1 for the initial $z$ field run with $\beta_z = 5000$, 
and is modestly greater for the run using a larger computational 
domain and an initial $\beta_z = 800$. These values are consistent with 
the values (0.1 - 0.2) obtained by Hawley, Gammie \& Balbus (1995) from  
unstratified local simulations in the shearing-box geometry, 
at comparable initial $\beta_z = 3200$. 
As expected, the fields are significantly weaker in the simulation 
with an initial $\phi$ field and non-periodic boundary conditions 
in $z$. In this case we obtain a ratio of magnetic to thermal energy 
in the $1 - 2 \times 10^{-2}$ range. In all the runs, modest field amplification 
interior to the marginally stable orbit is observed.
The fields remain well below equipartition with the thermal energy, while 
the kinetic energy is of course orders of magnitude greater still.

Angular momentum transport in the disk is dominated by magnetic, rather than 
fluid, stresses. We also plot in Fig.~6 a measure of the magnetic stress, 
normalised to the gas pressure,
\begin{equation} 
 \alpha_{\rm mag} = {2 \over 3} \langle { {-B_r B_\phi} \over {4 \pi \rho c_s^2} } \rangle,
\label{eq_alpha}
\end{equation} 
which in the disk is just the magnetic contribution to the Shakura-Sunyaev 
$\alpha$ parameter. Typical values obtained are a few 
$\times 10^{-2}$ for the standard simulation, and somewhat less than $10^{-2}$ 
for the simulation with an initial magnetic field in the $\phi$ direction. From the 
figure, it can also be seen that there is continuing magnetic stress, and a 
rise in $\alpha_{\rm mag}$, within the plunging region, where the density 
drops rapidly. In these plots there is evidently nothing special about 
the location $r=r_{\rm ms}$, and the magnetic torque does not vanish there.
The existence of this stress inside $r_{\rm ms}$, which is one 
of the predictions of Krolik (1999), has also been seen in previous 
numerical simulations (Hawley 2000; Hawley \& Krolik 2000).

For the vertically stratified run, we have also checked for 
the presence of significant variations of $\alpha_{\rm mag}$, 
and other quantities, with $z$. Interior to $r_{\rm ms}$, 
the strength of the magnetic stress, normalised to the gas 
pressure, is found to be greater near the midplane of the 
disk than at larger $z$, but only modestly so. 

To test the effect of this ongoing stress on the dynamics of the 
flow inside the last stable orbit, we plot in Fig.~7 the specific angular momentum $l$
of the flow as a function of radius. To reduce fluctuations in $l$, we 
average the specific angular momentum over several timeslices taken 
from near the end of the simulations. For all the runs, 
$l(r)$ in the disk ($r > r_{\rm ms}$) is close to the Keplerian value 
for circular orbits in the pseudo-Newtonian potential, apart from a 
deviation close to $r_{\rm ms}$ where the radial density gradient 
is becoming large. At a zero-torque boundary, we 
expect that $dl/dr$ vanishes. There is some variation in 
$l$ within the marginally stable orbit, even in the averaged 
profiles, and weak indications of a small systematic 
decline towards smaller radii. This is consistent with the 
aforementioned presence of magnetic stress inside $r_{\rm ms}$.
However, it is clear from Fig.~7 that for both the unstratified 
and stratified simulations the specific angular momentum 
is close to flat within $r_{\rm ms}$. The 
magnetic stress that exists within $r_{\rm ms}$ does not appear 
to be sufficient to change $l$ significantly within the last 
stable orbit. In this (restricted) sense, the numerical results are  
broadly consistent with the standard, purely 
hydrodynamic, picture of thin disk accretion accretion onto 
black holes (Abramowicz \& Kato 1989), which assumes 
a zero-torque boundary condition at the last stable orbit.

The highest resolution of the runs discussed here is obtained 
in the vertically stratified simulation. For this run, we plot 
in Fig.~8 the specific angular momentum from 5 timeslices, 
evenly spaced between $t=150$ and $t=200$. Significant 
fluctuations in $l$ at any given radius are present. The 
influence of such fluctuations on the average structure 
of the disk outside $r_{\rm ms}$ is unclear, though we 
speculate that they could lead to significant changes 
in, for example, the radiative efficiency of the flow.
However, in none of the slices do we see evidence for the clear and 
continuing decline in $l$, interior to the marginally 
stable orbit, that was obtained in the  
simulations of Hawley (2000).

\section{Discussion}
In this paper, we have used MHD simulations to study the transition 
between a geometrically thin accretion disk, in which inflow is driven 
by the rate at which turbulence can transport angular momentum outwards, and the 
unstable plunging region interior to the marginally stable orbit. We find 
that in many respects the transition resembles that expected on the basis of previous 
analytic and one-dimensional numerical calculations 
(Paczynski \& Bisnovatyi-Kogan 1981; Muchotrzeb \& Paczynski 1982; 
Muchotrzeb 1983; Matsumoto et. al. 1984; Abramowicz \& 
Kato 1989). In particular, we find no evidence, at least in pseudo-Newtonian 
simulations at the current  
resolution, for the extremely strong growth in the 
importance of magnetic fields (relative to the thermal or 
rest-mass energy of the flow) and 
associated dynamical effects discussed by Krolik (1999). Indeed, in 
these thin disk simulations, the average specific angular momentum 
is close to flat at and within the last stable orbit. In this 
respect, the numerical results, at least for the time averaged 
state, appear to be consistent with disk models computed using 
a zero-torque boundary condition at the 
last stable orbit. This differs from the results of some
previous simulations (Hawley 2000; 
Hawley \& Krolik 2000), in which an unmistakeable 
decline in the specific angular momentum inside $r_{\rm ms}$ 
was obtained. Although the conditions at $r_{\rm ms}$ are 
expected to vary between thin and thick disks, the  
model computed by Hawley (2000) and Hawley \& Krolik (2000) 
is thin enough at $r_{\rm ms}$ that $l(r)$ is close to 
Keplerian. We speculate that differences in the 
equation of state, spatial domain, or numerical resolution could  
be responsible for the discrepancy. Further simulations 
investigating these possibilities are in progress.

Although we have not found any strong dynamical effects associated 
with magnetic fields interior to the last stable orbit, we do 
see evidence for the essential ingredient -- ongoing magnetic 
stresses in the plunging region -- of recent models that have 
questioned the validity of a zero-torque boundary condition for 
black hole accretion (Gammie 1999; Krolik 1999; Agol \& Krolik 2000). 
The differences between these simulations, and those 
presented by Hawley (2000) and Hawley \& Krolik (2000), are thus
quantitative rather than qualititive in nature.
We would therefore emphasize that further 
improvements in the simulations are required to 
investigate the behavior of the vertically stratified flows 
more robustly, to extend the simulations to test the 
evolution of cooler flows crossing the last stable orbit, and 
to explore the continuum between geometrically thin and thick 
disks.

The simulations reported here are non-relativistic, and cannot -- even 
crudely -- be extended to model any of the additional complexity that arises if 
the accretion flow is onto a Kerr black hole. Considerable 
progress has been made in the development of numerical methods for 
general relativistic hydrodynamics and magnetohydrodynamics (e.g. 
Font 2000), but these methods are probably not yet able to follow 
a turbulent disk flow transiting the last stable orbit. We note, 
however, that the general relativistic MHD calculations that have 
already been reported do show important differences with  
non-relativistic analogs (Koide, Shibata \& Kudoh 1999; Meier 1999; 
Koide et al. 2000), 
and this should be borne in mind as an important caveat to the 
results presented here. The dynamics of the inner disk could also be modified 
by magnetic fields exterior to the disk, either in a disk 
corona (Miller \& Stone 2000; Hawley \& Krolik 2000), or by fields linking the 
disk to a spinning black hole (Livio 1999; Blandford 1999).
Given the diverse range of variability observed in accreting 
black hole systems, further exploration of such ideas is 
clearly warranted.

\acknowledgements

We thank the developers of ZEUS and ZEUS-MP for making these codes 
available as community resources through the Laboratory for 
Computational Astrophysics, and Charles Gammie, Julian Krolik 
and Bohdan Paczynski
for valuable comments. CSR acknowledges support from Hubble 
Fellowship grant HF-01113.01-98A. This grant was awarded by the Space 
Telescope Institute, which is operated by the Association of Universities 
for Research in Astronomy, Inc., for NASA under contract NAS 5-26555. 
CSR also thanks support from the NSF under grants AST~9876887 and AST~9529170.
JC was supported by NASA/ATP grant NAGS5-7723.

\newpage

\begin{figure}
\plotone{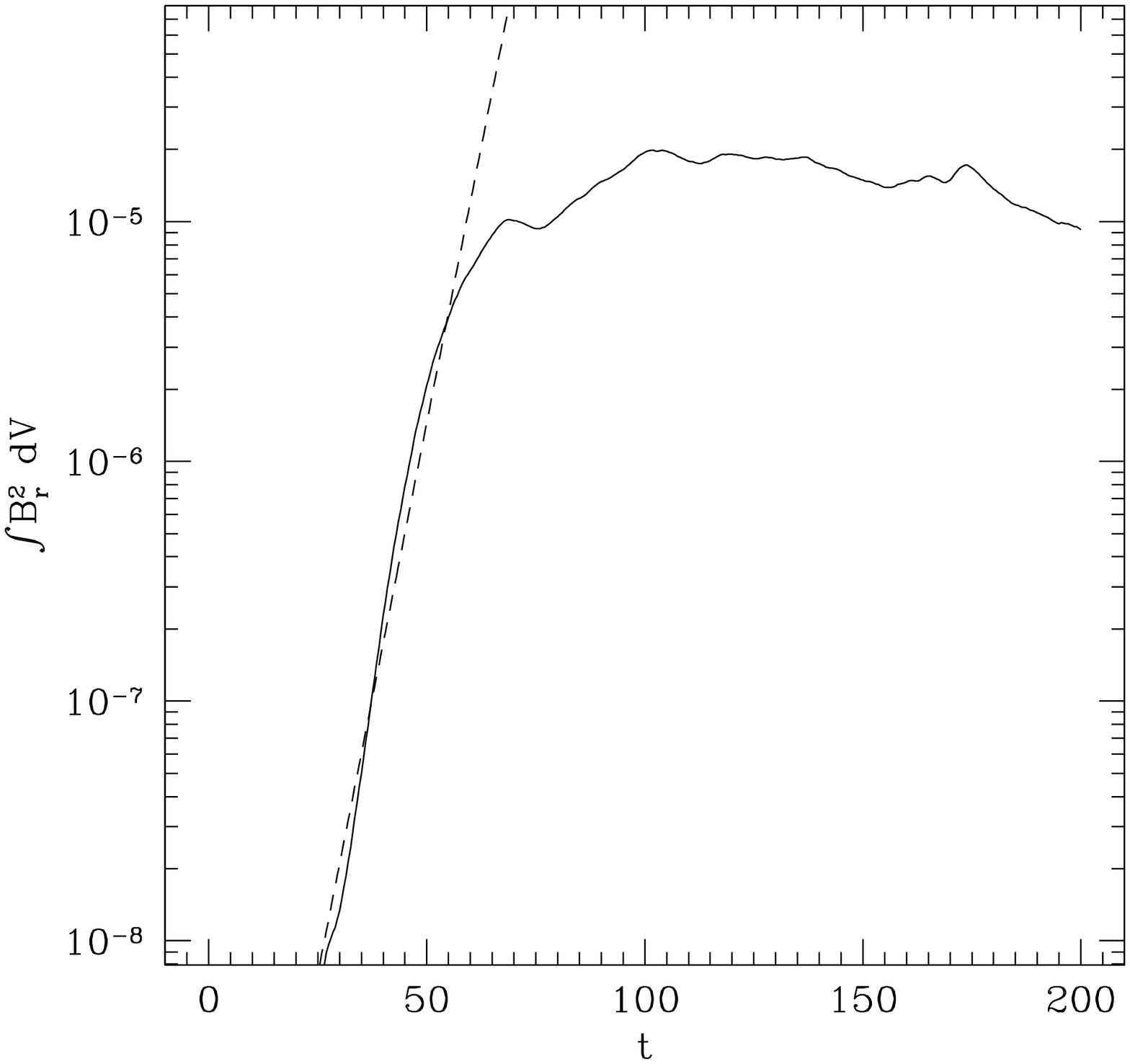}
\figcaption{Energy in the radial component of the magnetic field, 
		integrated over the simulation volume, as a 
	 	function of time. The units on the vertical axis 
		are arbitrary. The initial magnetic field 
		for this simulation was in the $z$ direction, 
		and the vertical boundary conditions were 
		periodic. The dashed line shows the expected growth 
		rate of the radial magnetic field energy density, 
		for the most unstable (largest $\Omega$) mode.} 
\end{figure}		
		
\begin{figure}
\plotone{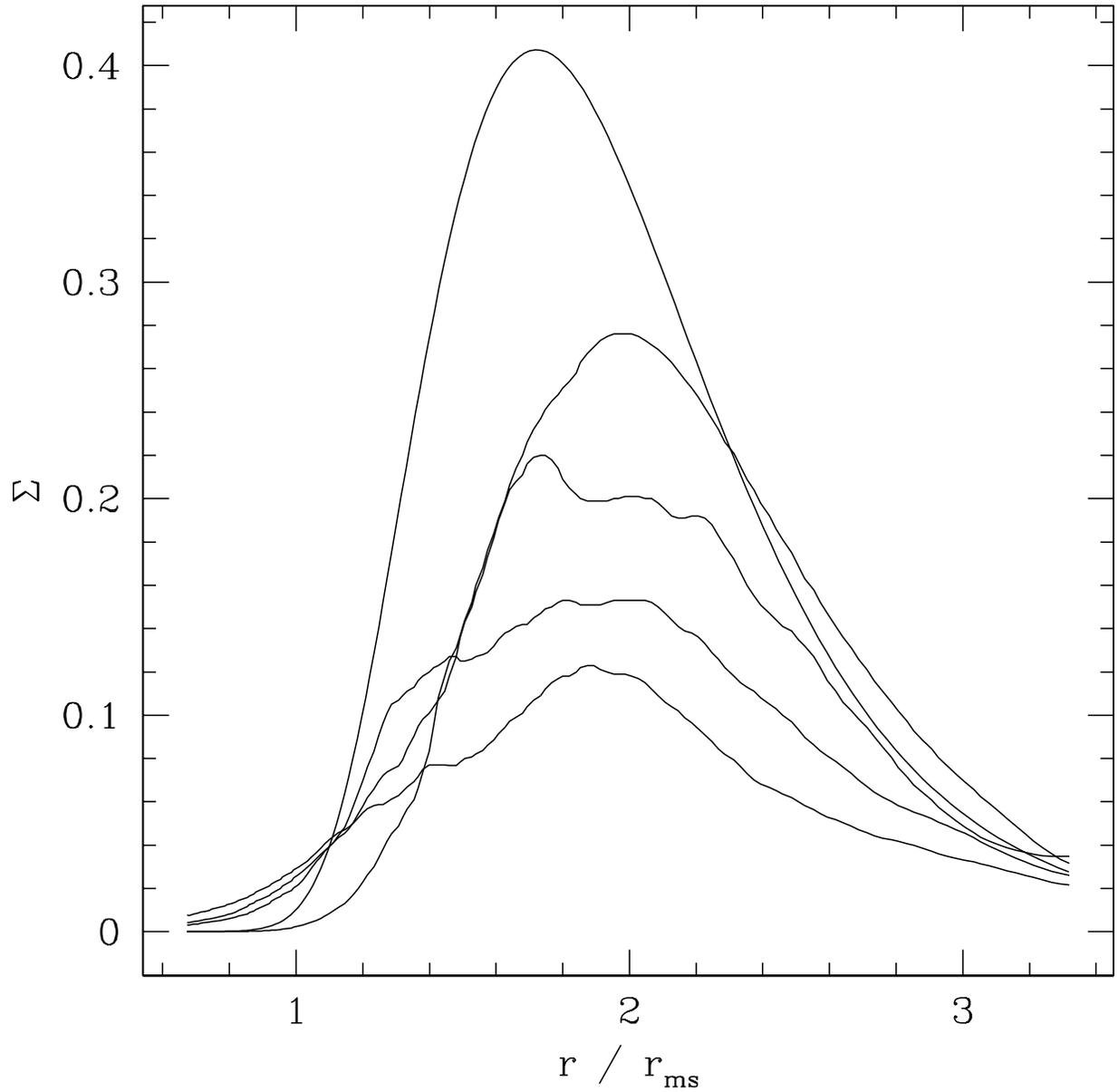}		
\figcaption{Evolution of the disk surface density profile in the simulation 
		with an initially vertical magnetic field geometry. From top down,
		the slices are plotted at $t=0$, $t=50$, $t=100$, $t=150$, 
		and $t=200$. More than half of the mass has been accreted 
		by the end of the simulation.}
\end{figure}		

\begin{figure}
\plotone{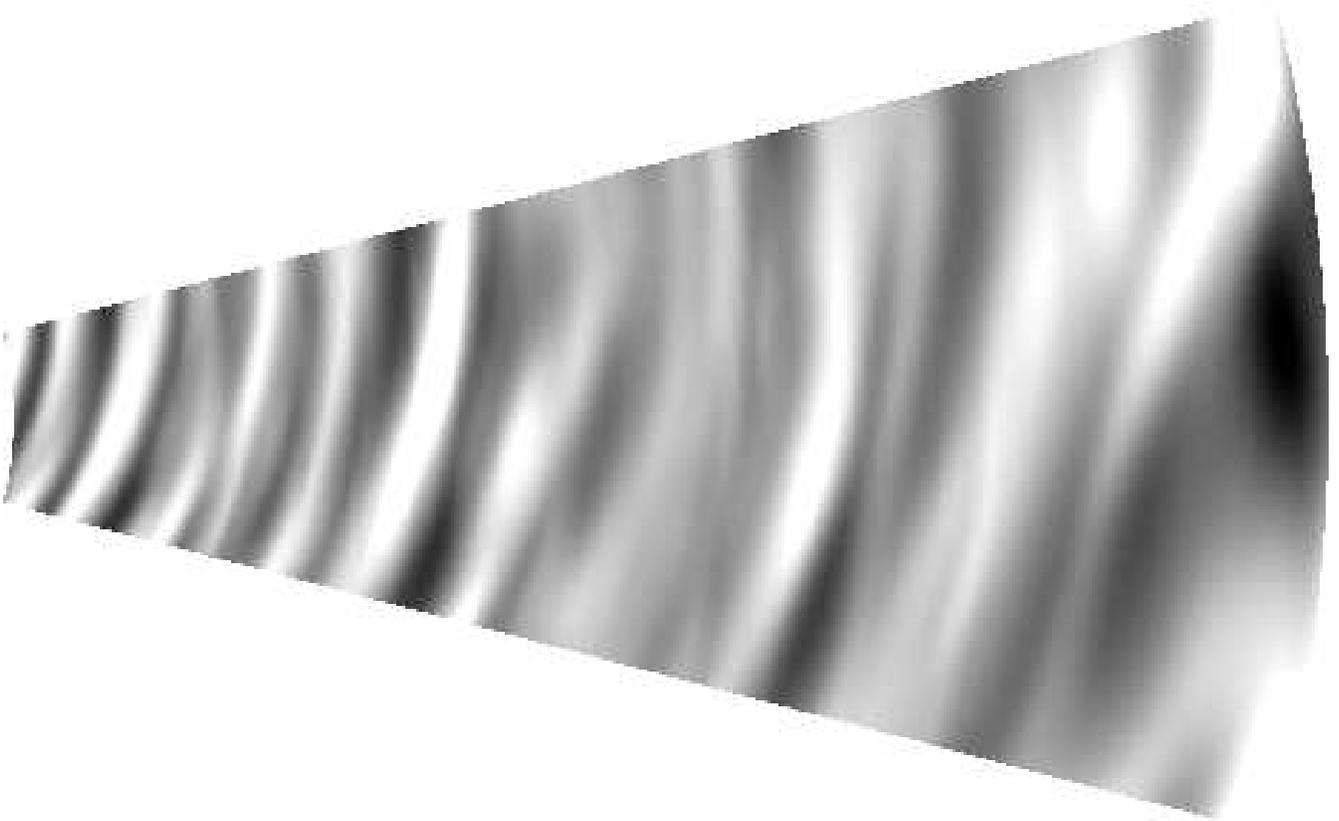}		
\figcaption{Image of the disk surface density fluctuations, $\Sigma(r,\phi) / 
           	\Sigma(r)$. The inner boundary is at $0.66 \ r_{\rm ms}$, the 
		outer boundary at $3.33 \ r_{\rm ms}$. The magnitude of the 
		fluctuations is at the $\sim$ 10\% level. No qualitative 
		changes are noticeable as the flow crosses the marginally 
		stable orbit.}	
\end{figure}			

\begin{figure}
\plotone{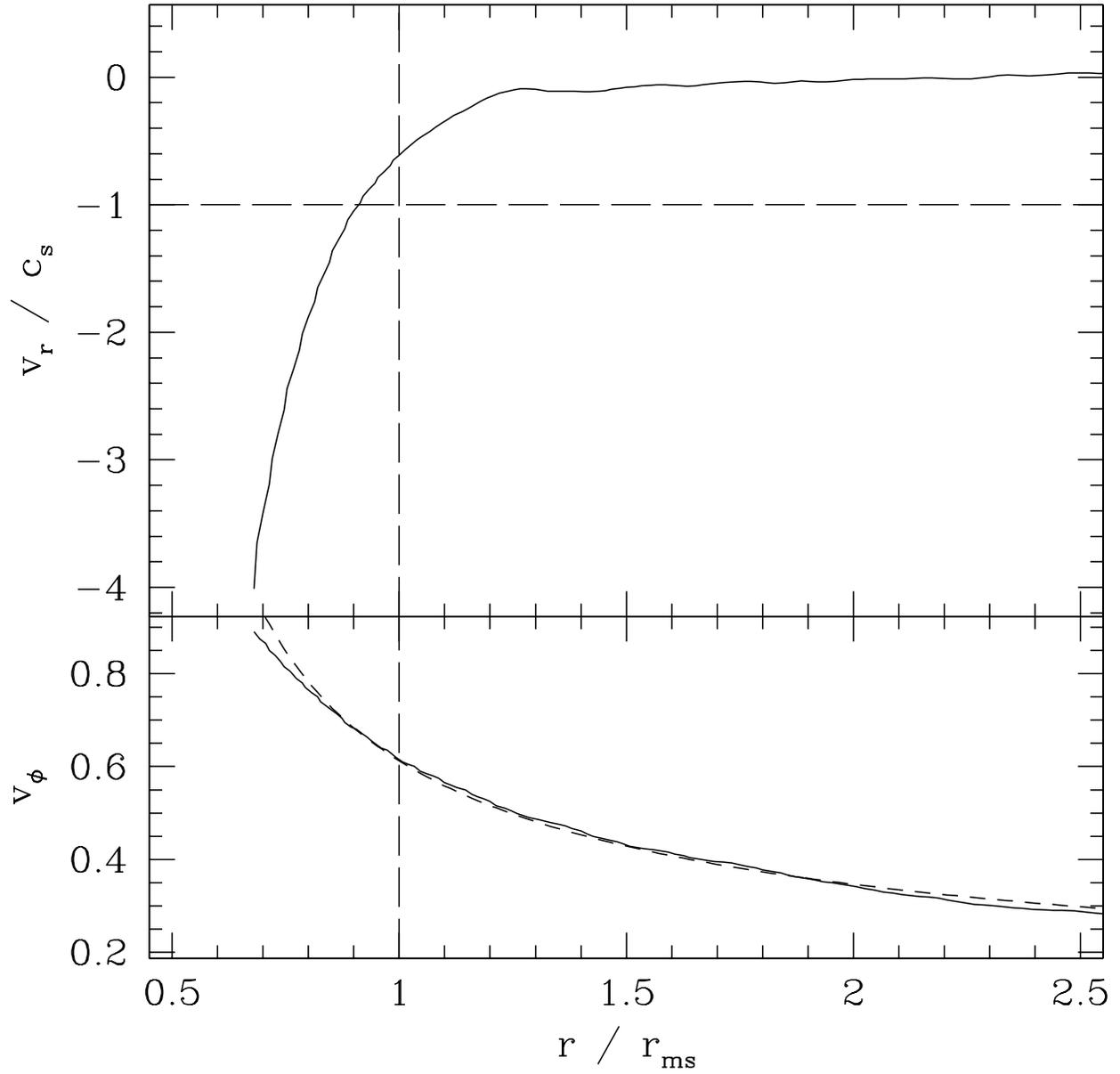}		
\figcaption{Radial and azimuthal velocity at $t=200$ from the standard simulation.
		The short dashed curve in the lower panel shows the Keplerian 
		velocity of circular orbits in the pseudo-Newtonian 
		potential, in units where $c=1$.}
\end{figure}

\begin{figure}
\plotone{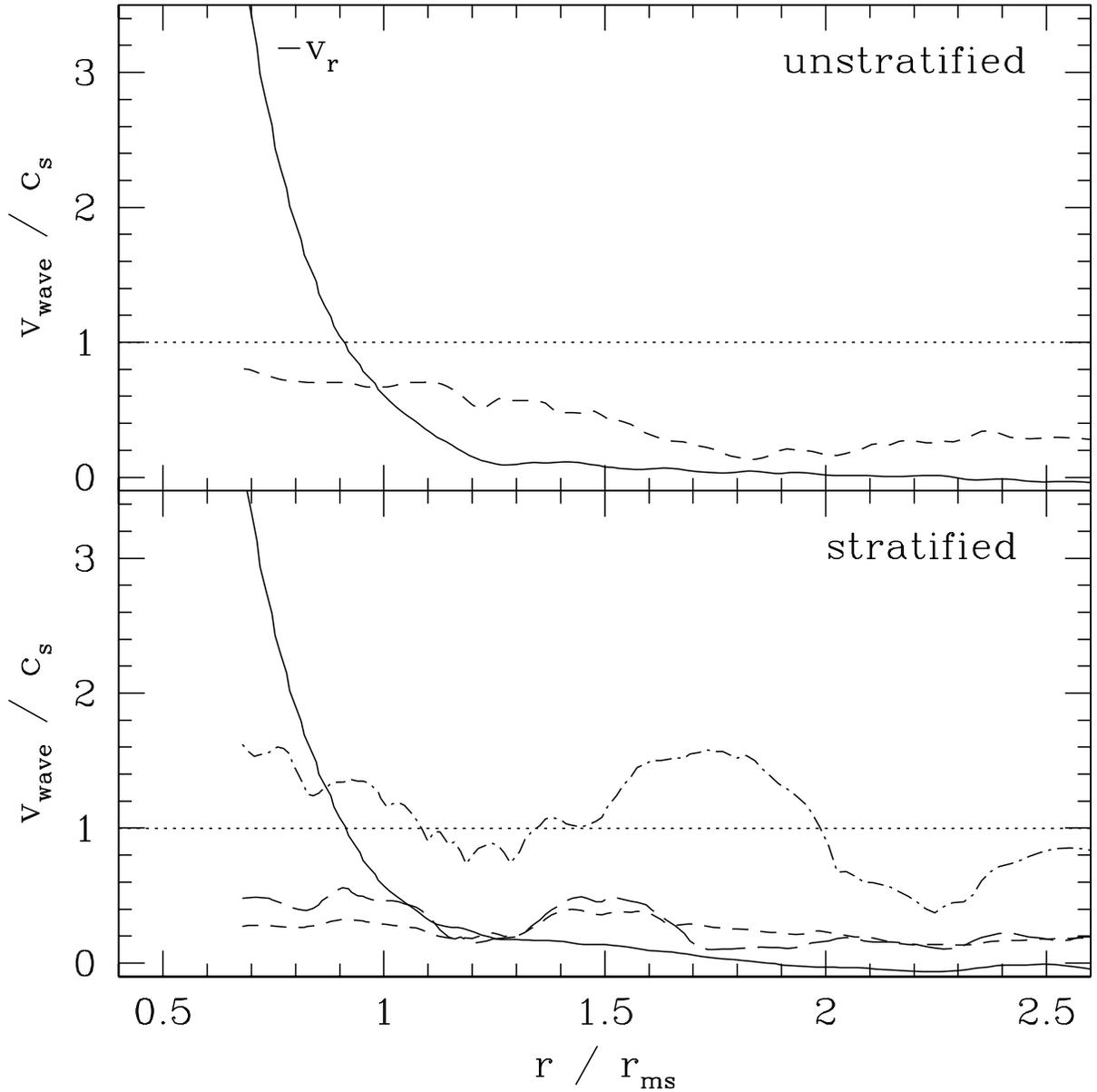}		
\figcaption{Wave propagation speed as a function of radius, 
		evaluated at $t=200$ from the standard simulation 
		with and without vertical stratification. In the 
		upper panel, for the unstratified run, 
		the horizontal dotted line shows the sound speed, 
		the short dashed line the mean Alfven speed as a 
		function of radius. 
		The negative of the radial velocity is plotted 
		as the solid curve. 
		The lower panel shows the same quantities plotted 
		for the stratified simulation, with the long dashed 
		line showing the mean Alfven speed in the flow 
		near the disk midplane ($\vert z \vert < 0.1$). 
		We also plot the {\em peak} Alfven speed at 
		each radius (dot-dashed line), which exceeds the mean value by 
		a substantial factor.}	
\end{figure}
\begin{figure}
\plotone{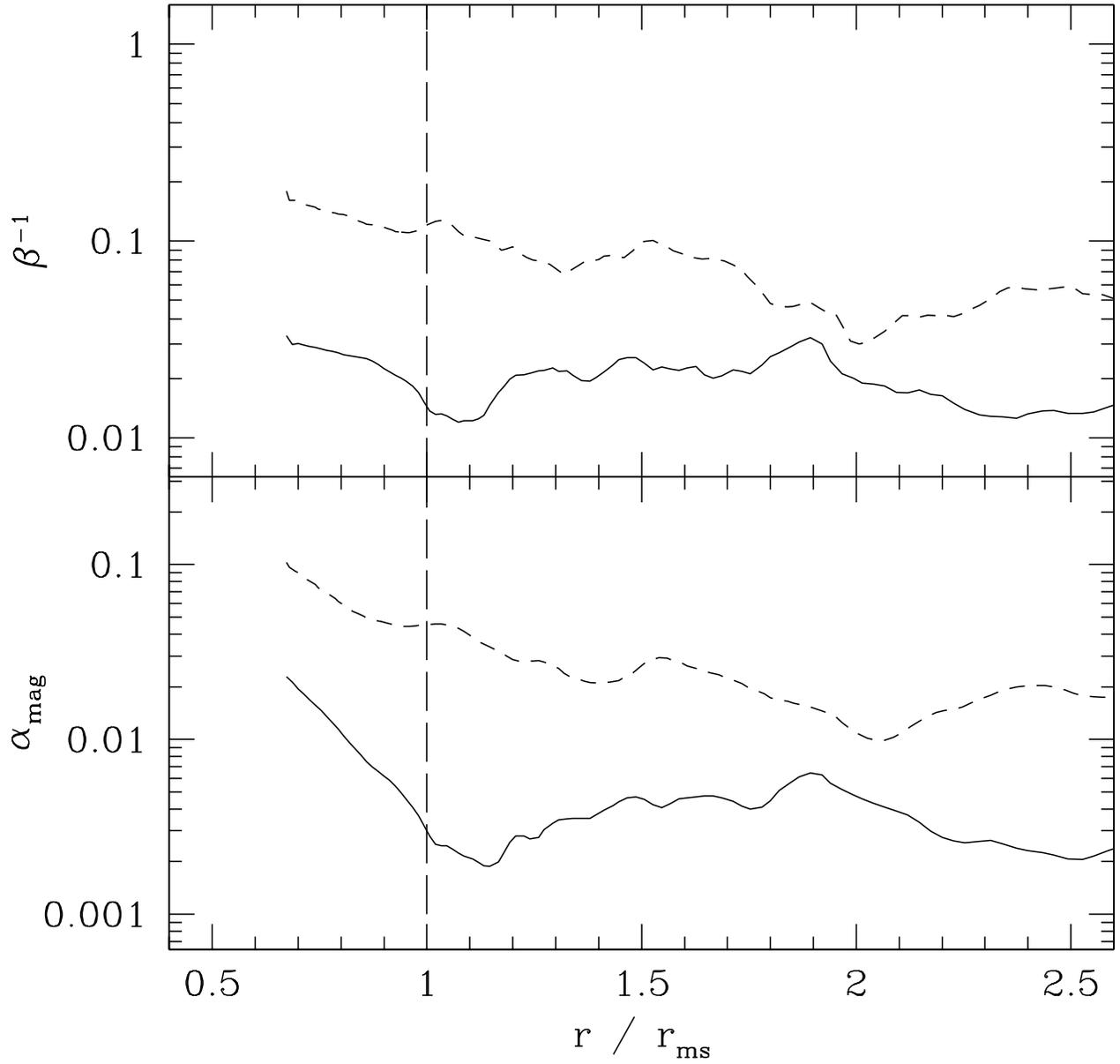}		
\figcaption{The upper panel shows the ratio of the magnetic energy to the thermal 
		energy, as a function of radius, for the standard simulation 
		with an initial $z$ field (dashed line), and for the run with 
		an initial $\phi$ field (solid line). In both cases, 
		several timeslices from near the end of the runs have been
		averaged together to reduce the magnitude of the fluctuations.
		The lower panel shows the effective Shakura-Sunyaev $\alpha$ 
		parameter derived from the magnetic torques.}	
\end{figure}		 

\begin{figure}
\plotone{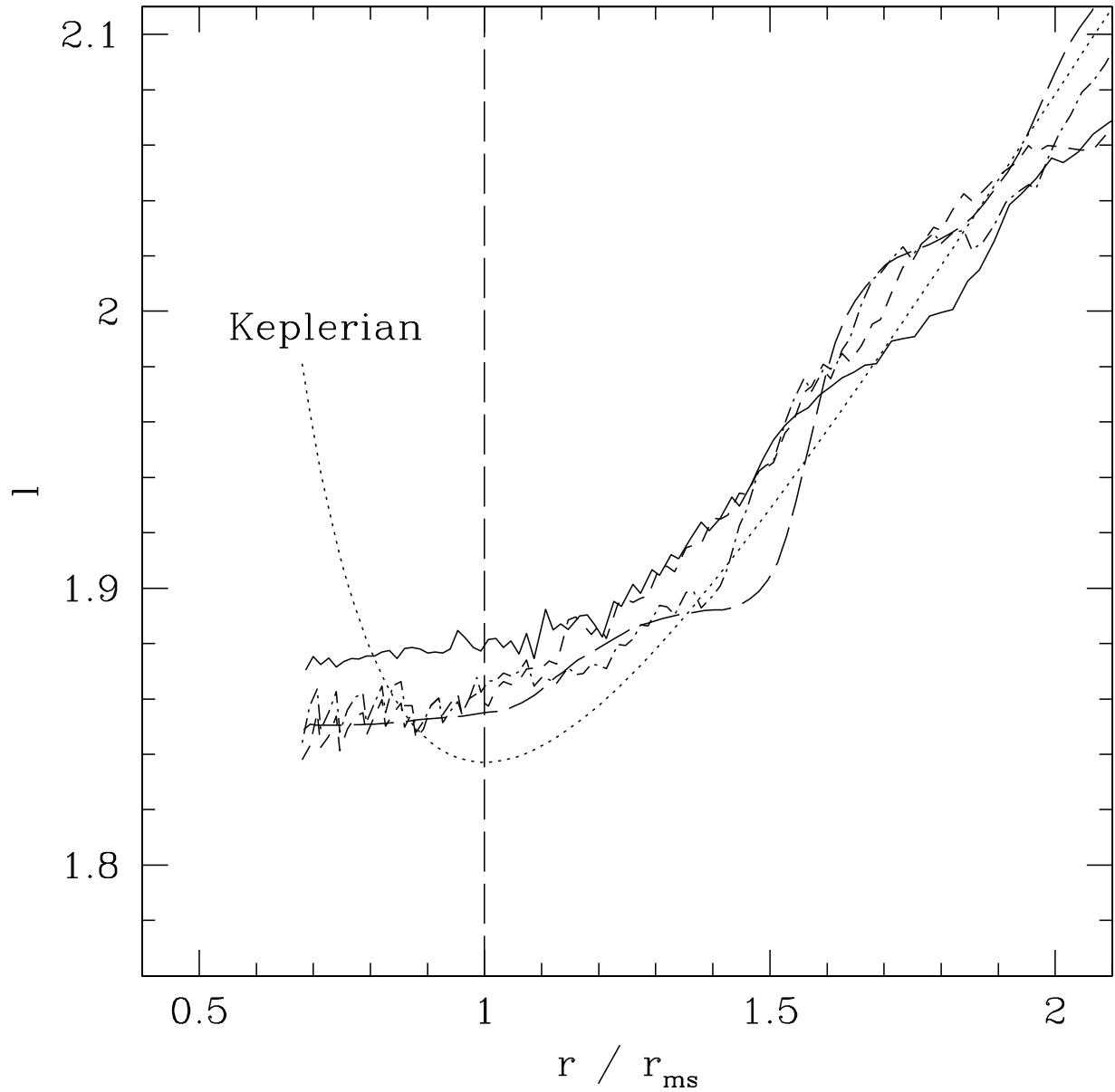}		
\figcaption{The specific angular momentum $l$ as a function of radius. The solid 
		line shows the result for the unstratified simulation with an initial $\phi$ 
		magnetic field, the short dashed and long dashed curves for unstratified 
		simulations
		with an initial $z$ field but different sound speeds and seed
		field strength. The dot-dashed line shows the result for the 
		vertically stratified simulation, evaluated in the 
		disk midplane.	
		Note that the simulation with reduced sound 
		speed (long dashed line) evolves more slowly, so that the 
		averaging of several timeslices has not eliminated substantial 
		fluctuations present at larger radius outside $r_{\rm ms}$.
		For all these models, $dl/dr$ is close to zero at $r_{\rm ms}$.
		The dotted curve shows the specific angular momentum of 
		circular orbits in the pseudo-Newtonian potential.}
\end{figure}

\begin{figure}
\plotone{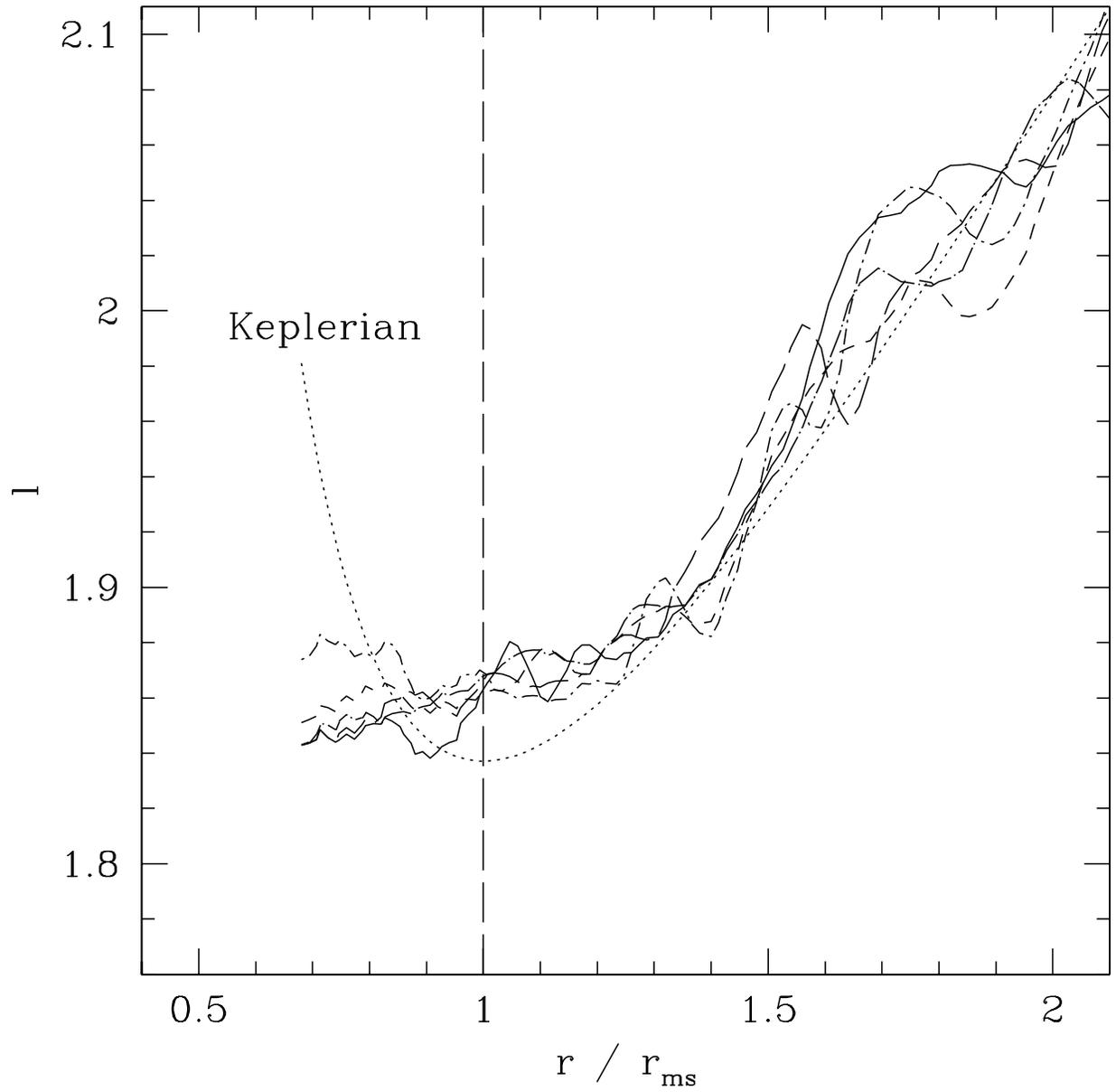}		
\figcaption{The specific angular momentum $l$ as a function of radius, plotted 
		from five timeslices of the stratified run. As in Fig.~7, 
		$l$ is evaluated in the disk midplane. 		
		The dotted curve shows the specific angular momentum of 
		circular orbits in the pseudo-Newtonian potential.}				                
\end{figure}

\end{document}